\documentstyle[preprint,prl,aps]{revtex}

\begin{document}
\draft
\title{Heterogeneity Induced Order in Globally Coupled Chaotic Systems}
\author{Tatsuo Shibata and Kunihiko Kaneko}
\address{Department of Pure and Applied Sciences, University of Tokyo,\\
		Komaba, Meguro-ku, Tokyo 153, Japan}
\date{\today}
\maketitle

\begin{abstract}
Collective behavior is studied in globally coupled maps with 
distributed nonlinearity.  It is shown that the heterogeneity
enhances regularity in the collective dynamics.   Low-dimensional
quasiperiodic motion is often found for the mean-field, even if each element
shows chaotic dynamics.  The mechanism of this order is
due to the formation of an internal bifurcation structure,
and the self-consistent dynamics between the structures and 
the mean-field.
\end{abstract}


\narrowtext

Dynamics of globally coupled systems has been extensively and intensively
studied [1-10].
Such problems naturally appear in physical and
biological systems.  Coupled Josephson junction array\cite{Wiesen}
and nonlinear optics with multi-mode excitation \cite{Roy,Arecchi}
give such examples,
while relevance to neural and cellular networks has been
discussed\cite{KK-rel}. Among others, study of globally coupled
chaotic systems has revealed novel concepts such as clustering,
chaotic itinerancy, partial order, and hidden coherence.

In particular, study of collective dynamics has  
gathered much attention [8-13].
An ensemble  of chaotic elements 
is found not to obey the law of large numbers 
and to form some kind of coherence \cite{KK-MF}.
Sometimes, quasiperiodic collective dynamics has been
found even if each element shows chaotic dynamics \cite{KK-MF,Pikov,KK-MF2}.

In these recent studies, elements are homogeneous,
in other words, identical elements are coupled.
However, in many systems elements are heterogeneous.
In Josephson junction array, each unit is
not identical.  In an optical system, the gain 
of each mode depends on its wavenumber.
In a biological system, each unit such as a neuron,
or a cell is heterogeneous.
So far the study of a coupled system with distributed parameters
is restricted to synchronization of a non-chaotic system \cite{Kura}.
Thus it is important to check how the notions 
constructed in globally coupled dynamical systems can be
applicable to a heterogeneous case. In the present Letter
we demonstrate that the collective order emerges in a heterogeneous
system through self-consistent dynamics between the mean-field and
an internal bifurcation structure.
Here we adopt a globally coupled map with a distributed parameter;
\begin{equation}
x_{n+1}(i) = (1-\epsilon)f_i(x_n(i))
+\frac{\epsilon}{N}\sum_{j=1}^{N}f_j(x_n(j)),
\end{equation}
\[i=1,2,3,\cdots,N\]
where  
$x_{n}(i)$ is the variable of the $i$th element
at discrete time $n$,
and $f_{i}(x(i))$ are the internal dynamics of each element.
For the functions we choose the logistic map
\[f_i(x) = 1-a(i) x^2,\] where
the parameter $a(i)$ for the nonlinearity is distributed between 
$[a_0 - \frac{\Delta a}{2},a_0 + \frac{\Delta a}{2}]$ 
as $a(i)=a_0 +\frac {\Delta a (2i-N)}{2N}$. We note that
the essentially same behavior is found when $a(i)$ is randomly
distributed in an interval or 
the coupling $\epsilon(i)$ is distributed instead of $a$.

When elements are identical with $\Delta a = 0$,
the present model reduces to a globally coupled map (GCM)
studied extensively.  In this case
there appears hidden coherence even when $x_{n}(i)$ shows chaotic oscillation 
and elements are totally desynchronized.
Indeed the variance of the mean field
\[h_n=\frac{1}{N}\sum_{j=1}^{N}f_j(x_n(j))\]
remains finite as the system size is increased,
in contrast with the expectation of the law of large numbers.
The mean square deviation (MSD) of the mean-field fluctuation, given
by $\langle (\delta h)^{2}\rangle 
= 
\langle (h - \langle h\rangle)^{2}\rangle$,
decreases proportional to $N^{-1}$ 
up to a crossover size $N_{c}$, but 
then remains constant with the further increase of $N$.
 
When
elements are not identical, one might expect that  hidden coherence 
would be lost and the law of large number could be recovered.
On the contrary, 
anomalous recovery of the MSD is found with the increase 
of the system size [8].
This suggests that some kind of order emerges 
in a heterogeneous system.  In the present Letter we clarify
the origin of such collective order.

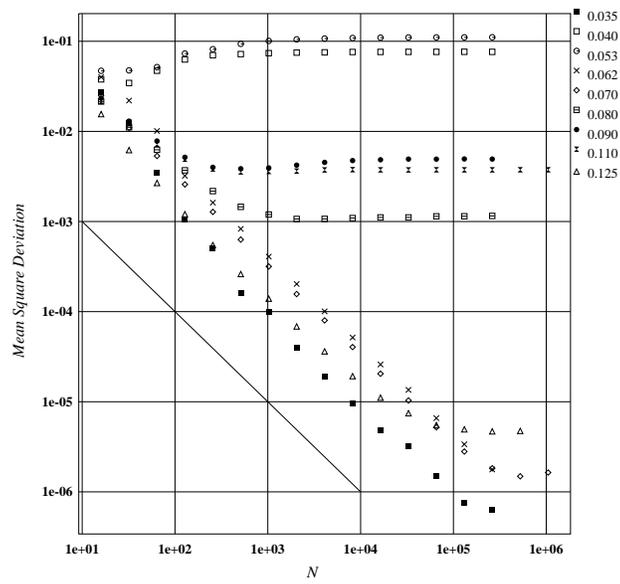
\begin{figure}
\caption{Mean square deviation (MSD) of the distribution of
the mean-field $h$ is plotted with the increase of the system size
$N$.$a=1.9, and \Delta a=0.05$, while the parameter $\epsilon$ is
shown at the right.}
\label{MSD_e}
\end{figure}

First we begin with the behavior of 
the mean field fluctuation in our system.
In Fig.\ref{MSD_e},
MSD is plotted with the increase of the system size $N$.
Roughly speaking
the MSD measures the amplitude of the mean field motion.
As the system size increases,
the MSD decreases up to a certain size and
then stays constant or increases to a certain constant.
This result implies the existence of 
some structure and coherence in the mean field dynamics.
The question we address here is their form and origin.

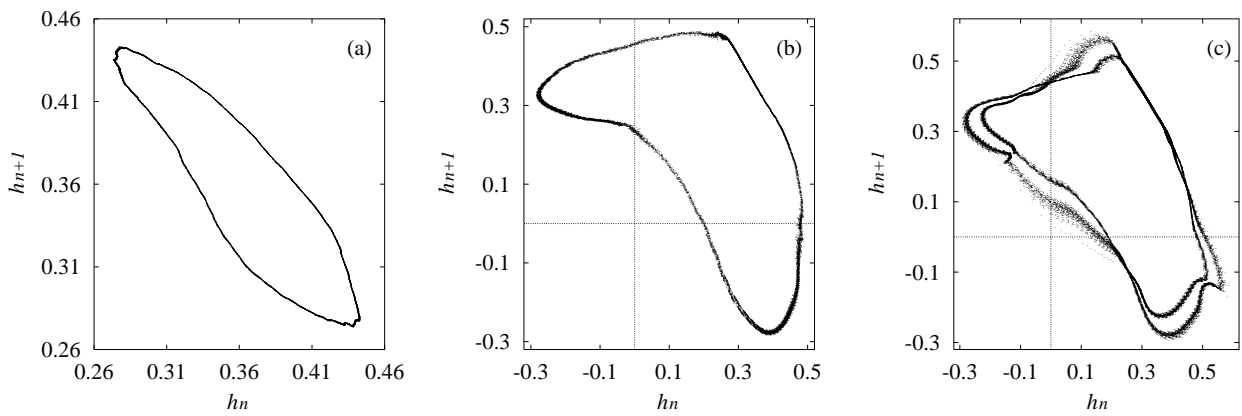
\begin{figure}
\caption{Return map of the mean-field $h$.
$a_{0}=1.9$ (a)$\Delta a=0.05$, $\epsilon=0.11$, $N=8\times 10^{6}$.
(b)$\Delta a=0.1$, $\epsilon=0.053$, $N=2^{18}$.
(c)$\Delta a=0.1$, $\epsilon=0.05$, $N=2^{21}$.}
\label{RMd05}
\end{figure}

Fig.\ref{RMd05}(a) gives an example of the return map of the mean field.
Here the width of scattered points along the
one-dimensional curve decreases with $N$.
Hence the figure  clearly shows
that the mean field dynamics is on a 2-dimensional torus.
The power spectrum of the mean field time series also supports that
the motion is quasi-periodic.
In this case, the oscillation has a frequency about 0.435.

From this, one can conclude that there appears collective order
in our system with the increase of the size. With the change of $a_{0}$,
$\Delta a$, or $\epsilon$, the mean-field dynamics shows the
bifurcation from torus to chaos accompanied by phase lockings.
Further bifurcation proceeds to much higher-dimensional chaos 
(while some structure is still kept).  
We have also observed the doubling of torus(Fig.2(b))
and other routes to chaos from quasiperiodic motions \cite{KK-book}.

There are two cases for the collective motion,
although for both cases
each element oscillates chaotically without mutual
synchronization.
In one case (given in Fig.2(a)), all Lyapunov exponents are positive
even if collective motion is quasiperiodic.
In the other case  (given in Fig.2(b)(c))some exponents are negative,
although most of them are positive.  
In this case, 
the origin of collective order is much easier to be detected.

It should be noted that such low-dimensional collective dynamics
is hardly observed in a globally coupled logistic map of identical parameters.
With a global scan of the parameter space, 
such low-dimensional collective dynamics is not observed
except for the trivial case, i.e.,
``direct product'' of periodic motions at a window, possible only for a 
narrow parameter region with a very small coupling
(e.g., $a \approx 1.8, \epsilon \approx .01$
it is 3-clustered motion).
Thus the heterogeneity in the parameter is essential to form the
low-dimensional collective dynamics.

Hereafter we show how this heterogeneity-induced order
is possible (mainly for the case with some negative Lyapunov exponents).
An example of the return map is given in Fig.2(b).
The scenario to be presented consists of two parts.  First, we demonstrate 
the formation of internal bifurcation structure, made possible
by the distribution of parameters, which leads 
to the self-consistent relation between each dynamics 
and the mean-field.
Second, it is shown that the
self-consistent dynamics is formed between the motion of the internal
bifurcation structure and the mean-field dynamics.

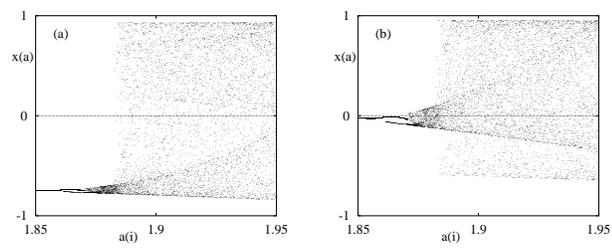
\begin{figure}
\caption{Internal Bifurcation Diagram.
$x_{n}(i)$ is plotted versus $a(i)$.
Here the mean-field is locked to period-3.
$a_{0}=1.9$, $\Delta a=0.1$, $\epsilon=0.055$.
Time step 5000 (a), and 5001 (b).
At the next iterate, 
the coherent structure of $x_{n}(a)$ for $a<1.88$ moves to $x\approx 1$, 
while another iterate leads to the structure of Fig.3(a)}
\label{DIAGRAM_a}
\end{figure}

First we study the formation of the internal bifurcation structure.
In our system 
nonlinear parameters are distributed over elements.
Dynamics of the $i$th element depends on
the parameter $a(i)$.
Hence it is relevant to draw the motion versus
the parameter $a$. 
Fig.\ref{DIAGRAM_a} gives snapshot patterns of $x_{n}(a)$
for the period-3 locking in the mean-field.
It looks like an ordinary bifurcation diagram 
plotted against the change of external parameters, 
but the patterns of Fig.\ref{DIAGRAM_a} are just 
snapshot representations of one system consisting of
$N$ elements, 
which is why we call the structure as internal bifurcation.

With the increase of $a(i)$,
tangent bifurcation, period doubling, and crisis are observed 
in the snapshot pattern,
if it is viewed as the transition of attractor 
with the change of control parameters.
If the mean-field were an external parameter 
for each element independent each other,
such viewpoint would hold. The mean-field, in our case, is
organized self-consistently from each element dynamics, where
the heterogeneity is the origin of the internal bifurcation.
The organization of the low-dimensional bifurcation
structures from a high-dimensional system
is a key concept for the collective dynamics
as will be discussed.

In Fig.\ref{DIAGRAM_a} elements
are almost synchronized for $1.85<a(i)<1.887$.
The motion of these elements is almost period-3.
Indeed, this motion comes from the
window in the single logistic map, which
plays an important role
to understand the mechanism of collective behavior.
Since the period-3 window is most prominent in the logistic map,
we explain the mechanism for this case, although
the explanation can be applied generally to other windows.

In the logistic map $f(x)=1-ax^2$, 
the period-3 window
appears through a tangent bifurcation of 
the 3rd iterate of the map $y=f(f(f(x)))$.
In this case, the map is tangential to $y=x$ at
3 points corresponding to the periodic points.
On the other hand,
for the mapping $f_{\delta_n}(x)=1-ax^2+\delta_n$
with a time-dependent parameter $\delta_n$,
the bifurcation usually occurs 
when the 3rd iterate
$y=f_{\delta 3}(f_{\delta 2}(f_{\delta 1}(x)))$
is tangential  to $y=x$ 
only at one point,
unless there is certain restriction to the
external field $\delta$.  In other words one
specific phase of the period-3 oscillation is selected
in accordance with the external parameter.

In Fig.\ref{DIAGRAM_a}, the
tangent bifurcation happens only at one point around $a\approx 1.85$.
As the parameter $a$ is larger,
the period doubling bifurcation and consequently 
crisis occurs
beyond which elements fall into a fully chaotic state.
Thus synchronized
and fully desynchronized motions
coexist.

This internal structure 
forms the mean-field as the ``input'' to each element self-consistently.
Period-3 clustered motion and fully desynchronized motion 
form the mean-field locked to period-3.
On the other hand, this period-3 ``input'' to each element
forms the internal structure mentioned above.
This self-consistent formation of clustered motion leads to
the simplest form of the collective order.
We note that such self-consistency is not attained in a homogeneous
system, except for a trivial case where
all elements are synchronized as a motion of a periodic-window.

When the coupling strength $\epsilon$ is smaller, another tangent
bifurcation occurs.  This bifurcation occurs after the complete
formation of the period-3 clustered motion mentioned above.  As the
2nd clustered motion is formed, the mean-field is varied, which
changes the internal bifurcation structure.  Then the period-3 locking
in the mean-field collapses, when we need the second scenario for the
self-consistent dynamics between the mean-field and the internal
bifurcation structure.

The scenario is summarized as follows: A coherent cluster is formed
for some parameter values of $a(i)$, by using a window structure in
the internal bifurcation, as is discussed.  Then the number of
elements belonging to this cluster increases, which leads to a
long-term change of the mean-field.  This, then, destabilizes the
cluster, but stabilizes another cluster with a different phase of
oscillation.  The latter cluster again gathers elements, which then
changes the mean-field in the opposite direction.  Repeating this
process, a long-term quasiperiodic oscillation is self-consistently
formed.

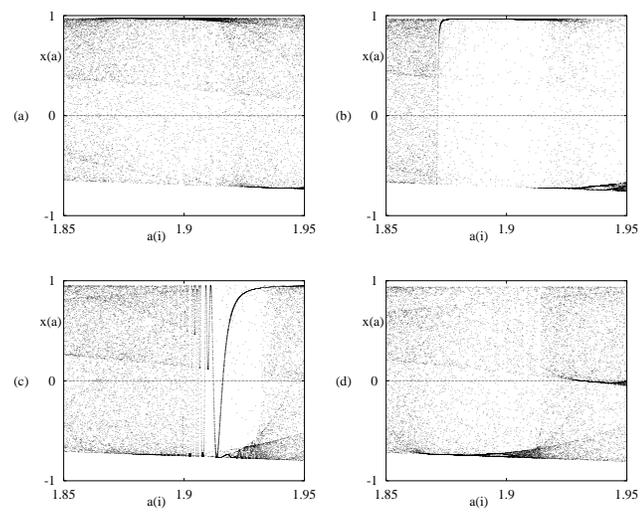
\begin{figure}
\noindent
\caption{Dynamics of the internal bifurcation structure.
$a_{0}=1.9$, $\Delta a=0.1$, $\epsilon=0.053$, 
at time step 5000 (a), 5081 (b), 5126 (c), 5271(d).}
\label{DIAGRAM_c}
\end{figure}

As the simplest example, we discuss
the quasi-periodic case given by  Fig.2(c).
Fig.\ref{DIAGRAM_c} shows the snapshots of $x(a)$ 
corresponding to the case of Fig.2(c).

To see the above scenario,
we study the change of internal structure of $x(a)$ at $3n$ step.
The process through Fig.\ref{DIAGRAM_c} is summarized as follows.
Two-clustered motion is formed,
although it does not last stably. 
The 1st cluster at $x=1$ (denoted as $c+$)
breaks down near $a=1.917$ by crisis,
and the elements whose $a(i)$'s are larger than this value
leave the cluster.  
On the other hand, after the formation of the 1st cluster,
the tangent bifurcation near $x=-0.7$ and $a=1.92$ occurs.
It forms the 2nd cluster (denoted as $c-$), 
which attracts elements (Fig. \ref{DIAGRAM_c}(a)), while
the 1st cluster at $x=1$ starts to collapse 
from smaller values of $a$ successively.
With this process, the 2nd cluster grows from lager $a$ to smaller
(Fig.\ref{DIAGRAM_c}(b)(c)). 
With the complete collapse of the 1st cluster, 
the 3rd cluster is formed at $x=0$ (denoted as $c0$) (Fig.\ref{DIAGRAM_c}(d)). 

The above process, taking about 170 time steps,
repeats successively changing the roles of
the three clusters $c+$, $c0$, and $c-$.
This repeated collapse and formation of the three clusters
is the origin of the quasi-periodic motion in the mean field.

To see our feedback scenario here, we need to
clarify (a) how the internal structure determines the mean-field
and (b) how the mean-field modifies the stability of clusters.
This is carried out by analyzing the change of 
\begin{equation}
x_{3n}=F_{h_{3n-1}}(F_{h_{3n-2}}(F_{h_{3n-3}}(x))),\label{LOGI_H3}
\end{equation}
which is the  3rd iterate of 
$F_{h_{n}}(x)=(1-\epsilon)(1-ax^2)+\epsilon h_{n}$,
where $h_{n}$ is the mean field as an external parameter 
for each element.

Let us consider the relationship 
between $c0$, $c+$, $c-$-clusters
and $h_{3n}$, $h_{3n-1}$, $h_{3n-2}$.
The step (a) is rather simple.  When, for example, the cluster $c+$ 
grows, $\frac{1}{N}\sum_{i=1}^{N}x_{3n}(i)$ is increased. 
Then $h_{3n-1}$ is increased, since 
\[h_{n-1} = \frac{1}{N}\sum_{i=1}^{N}f_i(x_{n-1}(i))
= \frac{1}{N}\sum_{i=1}^{N}x_{n}(i).\]
By a cyclic rotation of $c0$, $c+$, $c-$,
with the mapping from 3n to 3n+1 etc, other relationships 
between $h_{3n}$ and the cluster structure are obtained.

On the other hand,
a straightforward calculation of eq.(2) tells us that 
the (in)stability 
of $c_{+}$, $c_{0}$ or $c_{-}$ cluster
mainly depends on $h_{3n-1}$, $h_{3n-3}$ or $h_{3n-2}$ respectively. 
Each cluster is stabilized as
$h_{3n-1}$, $h_{3n-3}$ or $h_{3n-2}$ gets larger respectively.
Thus the process (b) is obtained.
From (a) and (b),
the stability 
of each cluster is mainly governed by
the change of the distribution 
of $x_{3n}$, $x_{3n-2}$ and $x_{3n-1}$
respectively.

Let us reconsider the above scenario in more detail.
After the formation of the $c_{+}$-cluster at $3n$ step,
the $c_{-}$-cluster starts to be stabilized from lager $a$.
Then the elements that left the $c_{+}$
by the crisis at $a=1.917$ are absorbed by $c_{-}$.
This formation of the $c_{-}$ cluster makes the mean-filed 
$h_{3n-1}$ to decrease(Fig.4(a)).

On the other hand,
this decrease of the mean-field modifies the stability of the $c_{+}$-cluster;
the tangent bifurcation point of $c_{+}$ moves to larger $a$.
Then the $c_{+}$ cluster is destabilized from smaller $a$
till it collapses(Fig.4(b)(c)).

Corresponding to the collapse of $c_{+}$-cluster at $3n$ step,
$c_{-}$ cluster starts to collapse at $3n-2$ step.
According to this, $h_{3n-3}$ starts to increase, which 
stabilizes the $c_{0}$ cluster at $3n$ step.
Now $h_{3n-1}$ starts to increase(Fig.4(d)), and so forth.

We note that the above feedback process between the mean-field and
internal bifurcation structure is possible, since the value of $a$ is
non-identical.  The role of elements is differentiated as to the
synchronization and desynchronization, which temporally changes as in
the case for chaotic itinerancy\cite{KK-GCM,CI1,CI2}.  We also note
that a slow modulation of the mean-field dynamics is formed by the
feedback.  This separation of time scales is necessary to have a
low-dimensional collective order; otherwise high-dimensional chaotic
dynamics remains in the mean-field as in the hidden coherence in GCM
\cite{KK-MF}.

Although we have explained the above scenario for the period-3 window
case due to its simplicity, this mechanism is generally applied
to our system, since each (logistic) dynamics contains a variety of
windows and bifurcations.   For example, we have seen the
change of the synchronization and internal bifurcation
structures for the parameters for Fig.2(a), where
all Lyapunov exponents are positive and clear windows are not visible.

To sum up, we have shown the formation of low-dimensional
collective dynamics in a coupled chaotic system with heterogeneity.
The mechanism of the formation is due to the
internal bifurcation structure afforded by heterogeneity, 
and the self-consistent feedback dynamics between the mean-field
and synchronization of elements.  We note that this
mechanism is expected to be quite general, as long as each local dynamics
allows for bifurcations with the change of some parameters,
distributed by elements.  Thus
our scenario for the collective order can be observed
in coupled systems such as Josephson junction arrays, and
multi-mode laser systems, as well as in biological networks.

The authors would like to thank N. Nakagawa for useful discussions.
This work is partially supported by Grant-in-Aids for Scientific
Research from the Ministry of Education, Science, and Culture
of Japan.

\end{document}